\documentclass[10pt,conference]{IEEEtran}

\usepackage{epsfig,rotating,setspace,latexsym,amsmath,epsf,amssymb,amsfonts,bm,theorem,cite,authblk, bbm,color, multirow, caption, subcaption, booktabs, xcolor}
\usepackage{makecell}

\IEEEoverridecommandlockouts
\allowdisplaybreaks

\title{Multistatic ISAC with Deep Learning-Based Clutter Suppression for 5G and Beyond}

\author{Yalin E. Sagduyu, Tugba Erpek, and Sastry Kompella}
\affil{\normalsize Nexcepta, Gaithersburg, MD, USA}

\date{}
\begin{document}

\bstctlcite{IEEEexample:BSTcontrol}

\maketitle
\thispagestyle{empty}
\begin{abstract}
Integrated sensing and communications (ISAC) systems using 5G and beyond signals of opportunity enable joint wireless sensing and communication using shared infrastructure and spectrum resources. While multistatic ISAC can increase sensing coverage with spatially distributed receivers, dense and dynamic clutter can significantly degrade bistatic range and Doppler estimation, leading to errors in target localization and velocity estimation. This paper presents a multistatic ISAC framework with clutter suppression for high-fidelity estimation of target position and velocity. Distributed clutter is modeled with randomly located scatterers with varying Doppler spreads and radar cross sections to capture dense and dynamic clutter effects. A broad range of conventional signal processing baselines are evaluated for clutter suppression, showing limited robustness under strong clutter conditions. To address this limitation, deep learning-based clutter suppression methods are introduced to improve robustness against dense and dynamic clutter by employing residual and attention-enhanced deep neural network architectures that directly enhance target responses from clutter-contaminated observations. These methods substantially outperform conventional approaches in target position and velocity estimation across diverse clutter conditions in multistatic ISAC.
\end{abstract}


\section{Introduction}
\emph{Integrated sensing and communications} (ISAC) is a key technology for next-generation wireless networks that repurposes communication signals for sensing using shared spectrum and infrastructure \cite{gonzalez2024integrated}. An important advantage of ISAC is the ability to exploit \emph{signals of opportunity}, such as commercial 5G NR signals, for target sensing and detection \cite{nataraja2024integrated,Wypich25,blandino2025detecting}. In particular,  5G NR \emph{positioning reference signals (PRS)} provide structured pilot patterns that can be effectively repurposed for passive sensing without dedicated radar transmitters and support persistent low-cost sensing over large areas \cite{wei20225g}. The ISAC capabilities are particularly important for \emph{mission-critical applications} such as intelligence, surveillance, and reconnaissance (ISR), border monitoring, and counter-unmanned aerial systems (counter-UAS), which require robust sensing, localization, and tracking in complex environments.

\emph{Monostatic} ISAC systems employ co-located transmitters and receivers,
which simplifies synchronization and signal processing but limits sensing
coverage and geometric diversity \cite{lu2022degrees, sagduyu2023joint, sagduyu2024will, iscout24, keskin2025fundamental}. In contrast, \emph{multistatic} ISAC systems
use spatially distributed receivers observing reflections from a common
transmitter, enabling improved spatial diversity, wider coverage,
enhanced target observability, and more accurate localization \cite{sagduyu2025multi, yang2025cooperative}. Despite these advantages, multistatic ISAC sensing remains challenging in dense \emph{clutter} environments, where delayed and Doppler-shifted clutter reflections can overlap with target returns in bistatic range-Doppler space. Since clutter characteristics can vary across spatially distributed receivers, clutter can significantly degrade target localization and velocity estimation performance \cite{luo2024integrated, liu2026clutter}. 

In this paper, we study multistatic ISAC with 5G NR PRS in clutter environments for bistatic range-Doppler estimation, two-dimensional (2D) target localization, and velocity estimation using spatially distributed receivers. A multistatic sensing framework based on cross-ambiguity-function (CAF) processing and nonlinear multilateration is adopted. The clutter is introduced using distributed scatterers with randomized locations, radar cross sections (RCSs), and Doppler spreads to capture diverse clutter conditions including highly dynamic clutter with significant Doppler overlap with target returns. We show that conventional signal processing-based \emph{clutter suppression} methods can only provide partial improvement and their effectiveness drops under dense or dynamic clutter conditions that degrade multistatic ISAC performance.

To improve robustness, we introduce \emph{deep learning} (DL)-based clutter suppression methods to directly enhance target responses from clutter-contaminated observations. The first method employs a residual U-Net architecture for target-enhanced residual estimation, while the second method introduces directional convolutional processing and attention-gated feature fusion to further improve clutter suppression, target enhancement, and multistatic sensing performance. We show that the proposed residual and attention-aware DL methods substantially outperform conventional clutter suppression approaches across a wide range of clutter conditions,
achieving significantly improved bistatic range-Doppler estimation, target localization, and velocity estimation performance.

 The remainder of the paper is organized as follows. Sec.~\ref{sec:model} presents signal and clutter models. Sec.~\ref{sec:detection} describes position and velocity estimation. Sec.~\ref{sec:conventional} discusses conventional clutter suppression. Sec.~\ref{sec:DL} presents DL-based clutter suppression. Section~V evaluates performance. Sec.~\ref{sec:conclusion} concludes the paper.

\begin{figure}[t!]
	\centering
	\includegraphics[width=0.58\columnwidth]{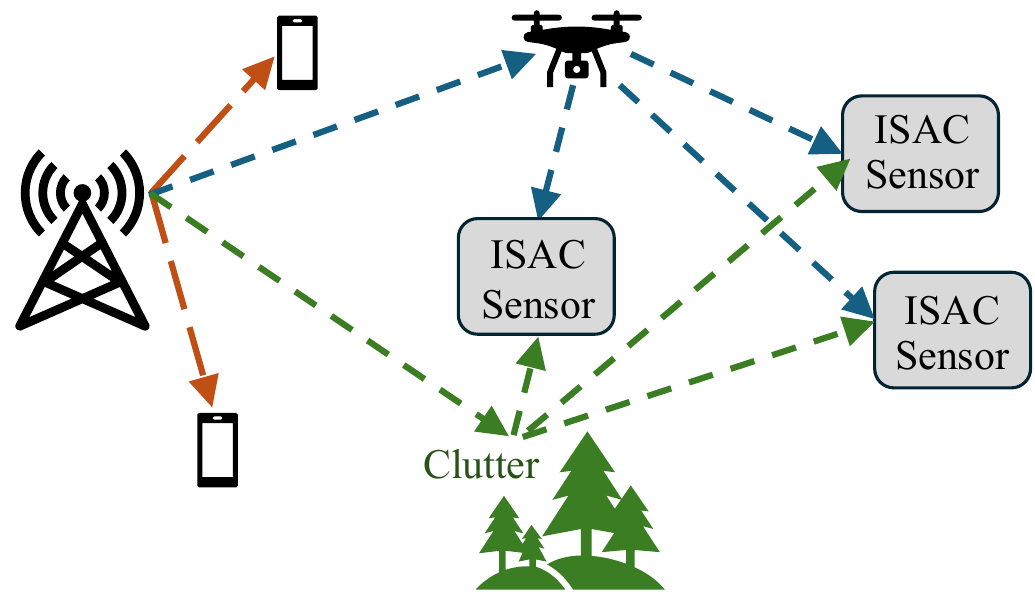}
	\caption{Multistatic ISAC in clutter environment.}
	\label{fig:system}
    \vspace{-0.3cm}
\end{figure}

\section{Multistatic ISAC Signal and Clutter Model \label{sec:model}}

We consider an illuminator (base station) that transmits 5G NR PRS and $M\!=\!3$ distributed receivers that cooperatively process the reflected signals for 2D sensing, as shown in Fig.~\ref{fig:system}. 5G numerology~0 is assumed as in  \cite{sagduyu2025multi} with subcarrier spacing
\(\Delta f \!= \!15~\mathrm{kHz}\), FFT size \(N\!=\!1024\), carrier frequency \(f_c\!=\!2.5~\mathrm{GHz}\), and sampling rate
\(f_s \!=\!15.36~\mathrm{MHz}\). The PRS spans
\(K_{\mathrm{RB}}\!=\!76\) resource blocks and follows the 3GPP TS~38.211 generation procedure using QPSK Gold sequences
\cite{3gpp.38.211}. Resource allocation uses a comb structure with spacing \(K_{\mathrm{comb}}\!\!=\!\!2\) and offset
\(k_{\mathrm{offset}}\!=\!0\), excluding the DC subcarrier. Each NR slot contains 14 OFDM symbols, including \(L_{\mathrm{PRS}}\!=\!2\) PRS symbols. Consecutive PRS transmissions
across multiple slots produce \(K\!\!=\!\!128\) coherent slow-time sensing observations. 

For receiver \(m \in \{1,2,3\}\), the bistatic
propagation length is
\begin{equation}
B_m(\mathbf{p})
=
\|\mathbf{p}-\mathbf{t}\|
+
\|\mathbf{p}-\mathbf{r}_m\|,
\end{equation}
where \(\mathbf{t}\) is the transmitter location,
\(\mathbf{r}_m\) is the receiver
\(m\) position, \(\mathbf{p}\) and 
\(\mathbf{v}\) are target position and velocity (all in \(\mathbb{R}^2\)).

The associated bistatic radial velocity at receiver \(m\) is
\begin{equation}
v_m(\mathbf{p},\mathbf{v})
=
\mathbf{v}^{\top}
\left(
\frac{\mathbf{p}-\mathbf{t}}
{\|\mathbf{p}-\mathbf{t}\|}
+
\frac{\mathbf{p}-\mathbf{r}_m}
{\|\mathbf{p}-\mathbf{r}_m\|}
\right),
\end{equation}
where $(\cdot)^{\top}$ denotes transpose. This represents the projection of the target velocity vector
$\mathbf{v}$ onto the bistatic observation direction associated with
receiver $m$. The resulting Doppler shift at receiver \(m\) is
\begin{equation}
f_{d,m}
=
\frac{f_c}{c}
v_m(\mathbf{p},\mathbf{v}),
\end{equation}
where \(c\) is the speed of light. The sensing region contains
\(N_c\)  randomly distributed clutter scatterers.
The \(q\)-th clutter scatterer with RCS \(\sigma_q^{(c)}\) is located at
\(\mathbf{c}_q\in\mathbb{R}^2\). The clutter RCS values are modeled as independent log-uniform random variables whose scale is controlled by the clutter-strength parameter $\alpha_c$, with larger $\alpha_c$ corresponding to stronger clutter returns. Each scatterer contributes a receiver-dependent Doppler component
\begin{equation}
f_{m,q}^{(c)}
\sim
\mathcal{N}(0,\sigma_{f,c}^2),
\end{equation}
where \(f_{m,q}^{(c)}\) is a zero-mean Gaussian random variable with variance \(\sigma_{f,c}^2\) and \(\sigma_{f,c}\) controls the clutter Doppler spread. Larger \(\sigma_{f,c}\) yields more dynamic clutter, while larger \(N_c\) yields denser clutter. The received signal at receiver \(m\) is the superposition of
the target echo, clutter reflections, and  noise:
\begin{align}
y_m[n]
&=
\alpha_m
s[n-\tau_m]
e^{j2\pi f_{d,m} nT_s}
\nonumber\\
&\quad+
\sum_{q=1}^{N_c}
\beta_{m,q}
s[n-\tau_{m,q}^{(c)}]
e^{j2\pi f_{m,q}^{(c)} nT_s}
+
w_m[n],
\end{align}
where \(s[n]\) denotes the transmitted PRS waveform,
\(T_s=1/f_s\) is the sampling interval,
\(\tau_m\) and \(\tau_{m,q}^{(c)}\) denote the target and clutter
propagation delays, respectively,
\(\alpha_m\) and \(\beta_{m,q}\) are their reflection
coefficients, and
\(w_m[n]\!\sim\!\mathcal{CN}(0,\sigma_n^2)\) is additive white
Gaussian noise. The reflection coefficients account for bistatic propagation loss,
antenna gains, carrier wavelength, and target and clutter RCS. With a narrowband approximation, the echo amplitude for a scatterer at location \(\mathbf{x}\) is modeled as
\begin{equation}
\gamma(\mathbf{x})
=
\frac{
\sqrt{P_T G_T G_R}\lambda \sqrt{\sigma}
}
{
(4\pi)^{3/2}
R_{\mathrm{tx}}
R_{\mathrm{rx}}
},
\end{equation}
where \(P_T\) is transmit power,
\(G_T\) and \(G_R\) are transmit and receive antenna gains,
\(\lambda\!=\!c/f_c\) is the carrier wavelength,
\(R_{\mathrm{tx}}\) and \(R_{\mathrm{rx}}\) are the transmitter-to-scatterer and scatterer-to-receiver distances, and
\(\sigma\) is the  RCS. 
Before CAF processing, the direct path is estimated and removed from the received signal.

\section{Target Localization and Velocity Estimation \label{sec:detection}}
Each sensing receiver correlates the received samples with the known PRS
reference sequence. Let \( 0 \leq k \leq K-1\) denote the slow-time PRS
observation index, \(t_k\) denote the midpoint time of the \(k\)-th PRS
symbol, and \(s_k[n]\), \(0 \leq n \leq N-1\), denote the associated
cyclic-prefix (CP)-free PRS reference symbol. For delay bin
\( 0 \leq d \leq D-1\), the correlation output is
\begin{equation}
C_m(k,d)
=
\sum_{n=0}^{N-1}
y_m[n_k^{\mathrm{cp}}+d+n]
s_k^*[n]
\end{equation}
at receiver \(m\), where \(n_k^{\mathrm{cp}}\) is the starting sample index of the
CP-free PRS region for observation \(k\) and \((\cdot)^*\) is the complex conjugate. The resulting matrix
\(\mathbf{C}_m \! \in \! \mathbb{C}^{K\times D}\) represents the slow-time
delay-response for clutter mitigation and Doppler analysis. Following clutter suppression (see Secs.~\ref{sec:conventional}-\ref{sec:DL}), coherent integration is applied over the slow-time dimension using a Doppler grid with
\(N_f\!=\!401\) frequency bins over \([-400,400]\)~Hz. Using a Hann
window \(W[k]\), the CAF  is
\begin{equation}
\label{eq:CAF}
\mathrm{CAF}_m(d,f)
=
\sum_{k=0}^{K-1}
W[k]
C_m(k,d)
e^{-j2\pi f t_k}
\end{equation}
for receiver \(m\) at Doppler frequency \(f\). The magnitude response
$|\mathrm{CAF}_m(d,f)|$
produces the bistatic range-Doppler map. Target detection uses two-dimensional cell-averaging constant false alarm rate (CA-CFAR) on the CAF magnitude response.
Three-point parabolic interpolation around the detected cell is 
applied in delay and Doppler dimensions to refine bistatic
range and radial velocity estimates 
\(\hat B_m\) and \(\hat v_m\). Target localization jointly fits the bistatic range
measurements from all receivers. Each bistatic measurement defines an
ellipse with foci located at the transmitter and the corresponding
receiver position. When
\begin{equation}
\epsilon_m(\mathbf{p})
=
\|\mathbf{p}-\mathbf{t}\|
+
\|\mathbf{p}-\mathbf{r}_m\|
-
\hat B_m
\end{equation}
is the bistatic residual associated with receiver \(m\), the estimated
target position is determined from solving
\begin{equation}
\hat{\mathbf{p}}
=
\arg\min_{\mathbf{p}}
\sum_{m=1}^{M}
\left( \epsilon_m(\mathbf{p}) \right)^2.
\end{equation}

Because the bistatic constraints are nonlinear, a trust-region
reflective least-squares solver with multiple randomized initialization
points is employed. The solution yielding the minimum aggregate residual
error is selected as the final estimate.

After estimating target position, the bistatic radial velocity
relations become linear in the Cartesian velocity vector. Define
\begin{equation}
\mathbf{u}_m
=
\frac{\hat{\mathbf{p}}-\mathbf{t}}
{\|\hat{\mathbf{p}}-\mathbf{t}\|}
+
\frac{\hat{\mathbf{p}}-\mathbf{r}_m}
{\|\hat{\mathbf{p}}-\mathbf{r}_m\|},
\end{equation}
as the bistatic observation direction for receiver \(m\).
Stacking these vectors forms the observation matrix
\(\mathbf{U}\in\mathbb{R}^{M\times2}\), while the estimated radial
velocities are collected into
\begin{equation}
\hat{\mathbf{v}}_{\mathrm{r}}
=
[\hat v_1,\ldots,\hat v_M]^\top.
\end{equation}

The Cartesian velocity is estimated using ridge-regularized least squares (with a small regularization parameter \(\varepsilon\!>\!0\))
\begin{equation}
\hat{\mathbf{v}}
=
\left(
\mathbf{U}^{\top}\mathbf{U}
+
\varepsilon\mathbf{I}
\right)^{-1}
\mathbf{U}^{\top}
\hat{\mathbf{v}}_{\mathrm{r}},
\end{equation}
where \(\mathbf{I}\) is the identity matrix. The target
speed is given by \(\|\hat{\mathbf{v}}\|\) and the heading angle is
computed using
\(\operatorname{atan2}(\hat v_y,\hat v_x)\).

\section{Baselines for Clutter Suppression \label{sec:conventional}}

We next compare a broad set of conventional clutter suppression methods operating on the slow-time delay correlation matrix $\mathbf{C}_m$. They suppress clutter using techniques like temporal differencing, subspace projection, Doppler-domain filtering, adaptive prediction, and smooth-component estimation. After clutter suppression, the resulting correlation matrices are used for CAF computation and  bistatic range-Doppler estimation.

\subsubsection{No Clutter Suppression (NCS)}
NCS skips clutter mitigation and computes the CAF from the correlation matrix
\(\mathbf{C}_m\).

\subsubsection{Current-Reference Subtraction (CUR) \label{sec:CUR}}

CUR serves as a coherent reference-based clutter suppression benchmark.
Let
\(\mathbf{C}_m^{\mathrm{cur}}\)
denote the current clutter-contaminated correlation matrix and
\(\mathbf{C}_m^{\mathrm{ref}}\)
denote the clutter-only reference correlation matrix for receiver \(m\).
A global clutter phase offset
\(\phi_m\)
is estimated from dominant clutter components in the reference and current observations. The current observation is phase-aligned prior to subtraction, yielding the clutter-suppressed correlation matrix
\begin{equation}
\mathbf{C}_m^{\mathrm{CUR}}
=
e^{-j\phi_m}
\mathbf{C}_m^{\mathrm{cur}}
-
\mathbf{C}_m^{\mathrm{ref}}.
\end{equation}
The clutter-suppressed matrix $\mathbf{C}_m^{\mathrm{CUR}}$ is used for CAF computation according to (\ref{eq:CAF}) to estimate bistatic range and Doppler.

\subsubsection{Space-Projection Adaptive Difference (SP-AD)}

SP-AD suppresses slowly varying clutter by differencing adjacent slow-time snapshots prior to CAF processing.

\subsubsection{Rank-One Slow-Time Projection (SP-R1)}

SP-R1 models clutter as a dominant rank-one slow-time component and removes the strongest singular mode before Doppler processing.

\subsubsection{DPSS Band-Stop Filtering (DPSS)}
Discrete prolate spheroidal sequences (DPSS) model clutter as a low-Doppler process and aim to suppress clutter by projecting the signal onto the orthogonal complement of a DPSS clutter subspace.

\subsubsection{Clutter Map Temporal Doppler Filtering (CMTD)}
CMTD suppresses near-zero-Doppler clutter using FFT-based slow-time notch filtering centered near zero frequency.

\subsubsection{Total Variation High-Pass Filtering (TVHP)}
TVHP models clutter as a temporally smooth component and removes smooth clutter responses using total-variation regularization.

\subsubsection{Rolling-Median High-Pass Filtering (RMHP)}

RMHP suppresses clutter using a sliding-window median filter along the slow-time dimension to remove slowly varying clutter responses while preserving localized target peaks.

\subsubsection{Wavelet High-Pass Filtering (WHP)}
WHP applies Haar wavelet decomposition along slow time and reconstructs the signal using high-frequency detail components.

\subsubsection{Adaptive Least Mean Squares (LMS) Residual Filtering (ALMS)}
ALMS applies adaptive LMS prediction along slow time and uses the prediction residual for clutter suppression.

\subsubsection{Polynomial High-Pass Filtering (POLYHP)}
POLYHP models clutter as a slowly varying low-order polynomial along slow time and removes the estimated polynomial component.

\section{Deep Learning-Based Clutter Suppression \label{sec:DL}}

In addition to signal processing-based clutter suppression methods, we introduce two supervised DL methods that operate directly on the slow-time delay correlation matrix
\(\mathbf{C}\in\mathbb{C}^{K\times D}\) obtained from matched filtering. The objective is to learn a nonlinear mapping that suppresses clutter while preserving target responses required for accurate bistatic range-Doppler estimation. Training data are generated from the multistatic ISAC signal and clutter model described previously. Each training realization contains random receiver geometries, target positions, target velocities, clutter densities, clutter RCSs, clutter Doppler spreads, and noise realizations. The training input is the clutter-contaminated matched-filter matrix
\begin{equation}
\mathbf{C}_{\mathrm{raw}}
=
\mathbf{C}_{\mathrm{target}}
+
\mathbf{C}_{\mathrm{clutter}}
+
\mathbf{N},
\end{equation}
where \(\mathbf{C}_{\mathrm{target}}\) and \(\mathbf{C}_{\mathrm{clutter}}\) denote the target-only
and clutter-only correlation
matrices, respectively, and
\(\mathbf{N}\) denotes the additive noise. The desired output is
\(\mathbf{C}_{\mathrm{target}}\). 

Next, we present two DL-based clutter suppression (DLCS) methods that estimate
$\mathbf{C}_{\mathrm{target}}$ from $\mathbf{C}_{\mathrm{raw}}$. The estimated
target-only correlation matrix is subsequently used in (\ref{eq:CAF}) in place
of $\mathbf{C}_m$ for range-Doppler estimation.

\subsection{Residual U-Net-based DLCS}

The first DL method, DLCS-1, employs a supervised residual U-Net architecture that directly estimates the target-only matched-filter response from clutter-corrupted observations.

\subsubsection{Input Representation} The input is given by

\begin{equation}
\mathbf{X}
=
[
\mathbf{C}_{\mathrm{raw}}^{(\mathrm{I})},
\mathbf{C}_{\mathrm{raw}}^{(\mathrm{Q})}
],
\end{equation}
where
\(\mathbf{C}_{\mathrm{raw}}^{(\mathrm{I})}\)
and
\(\mathbf{C}_{\mathrm{raw}}^{(\mathrm{Q})}\)
are the in-phase and quadrature components of the complex-valued correlation matrix, respectively, and are normalized using the root-mean-square (RMS) magnitude of the input tensor before being passed to the network.

\subsubsection{Network Architecture}

The network follows an encoder-decoder U-Net structure with skip connections. The encoder consists of repeated convolutional blocks that progressively extract hierarchical clutter and target features while reducing spatial resolution through max-pooling operations. Each convolutional block contains \(3\times3\) convolutional layers, batch normalization, and rectified linear unit (ReLU) activations.

The feature tensor \(\mathbf{F}^{(\ell)}\) at layer \(\ell\) is encoded as
\begin{equation}
\mathbf{F}^{(\ell+1)}
=
\mathrm{ReLU}
\left(
\mathrm{BN}
\left(
\mathrm{Conv}
(
\mathbf{F}^{(\ell)}
)
\right)
\right),
\end{equation}
where
\(\mathrm{Conv}(\cdot)\)
denotes convolution and
\(\mathrm{BN}(\cdot)\)
denotes batch normalization. The decoder progressively reconstructs a target-enhanced residual map using bilinear upsampling and skip-connected encoder features. Skip connections preserve fine spatial structures corresponding to localized target peaks in the range-Doppler domain. The final output layer is a \(1\times1\) convolution that produces a
two-channel estimate
\begin{equation}
\hat{\mathbf{Y}}
=
[
\hat{\mathbf{C}}_{\mathrm{target}}^{(\mathrm{I})},
\hat{\mathbf{C}}_{\mathrm{target}}^{(\mathrm{Q})}
],
\end{equation}
where
\(\hat{\mathbf{C}}_{\mathrm{target}}^{(\mathrm{I})}\)
and
\(\hat{\mathbf{C}}_{\mathrm{target}}^{(\mathrm{Q})}\)
denote the predicted in-phase and quadrature components of the target-only correlation matrix.

\subsubsection{Training Objective}

The network is trained using a residual reconstruction loss composed of
mean-squared error and sparsity regularization:
\begin{equation}
\mathcal{L}_{\mathrm{DLCS}}
=
\left\|
\hat{\mathbf{C}}_{\mathrm{target}}
-
\mathbf{C}_{\mathrm{target}}
\right\|_2^2
+
\lambda_1
\left\|
\hat{\mathbf{C}}_{\mathrm{target}}
\right\|_1,
\end{equation}
where
\(\lambda_1>0\)
controls sparsity regularization. The \(\ell_2\) term encourages accurate target reconstruction. The
\(\ell_1\) term promotes sparse target responses and suppresses residual
clutter artifacts.

\subsubsection{Optimization}

The network parameters are optimized using the AdamW optimizer with a learning rate
\(\eta_{\mathrm{DLCS}} = 10^{-3}\)
and a weight-decay regularization coefficient
\(\lambda_{\mathrm{wd}} = 10^{-5}\).
Mini-batch training is used with random clutter and
target realizations. Gradient clipping is applied to stabilize optimization:
\begin{equation}
\|\nabla_{\theta}\mathcal{L}\|_2
\leq
\gamma_{\max},
\end{equation}
where
\(\theta\)
denotes the trainable network parameters and the clipping threshold \(\gamma_{\max}\) is \(5\). After training, the  output \(\hat{\mathbf{C}}_{\mathrm{target}}\)
is processed in the CAF and range-Doppler estimation stages.

\subsection{Attention-Gated Anisotropic DLCS}

The second DL method, DLCS-2, extends the residual U-Net architecture through anisotropic convolutional processing and attention-gated skip connections designed specifically for clutter-dominated range-Doppler maps.

\subsubsection{Multi-Channel Input Representation}

Instead of using only real and imaginary channels, the network exploits
four complementary input representations:
\begin{equation}
\mathbf{X}
=
[
\mathbf{C}_{\mathrm{raw}}^{(\mathrm{I})},
\mathbf{C}_{\mathrm{raw}}^{(\mathrm{Q})},
|\mathbf{C}_{\mathrm{raw}}|,
\log(1+|\mathbf{C}_{\mathrm{raw}}|)
].
\end{equation}

The magnitude channel explicitly captures target energy distributions,
while the logarithmic channel compresses dynamic range and improves
representation of weak target responses in strong clutter environments.

\subsubsection{Anisotropic Convolutional Encoder}

The encoder uses anisotropic convolutional blocks that separately model slow-time and delay-domain structures. Each anisotropic block contains  standard \(3\times3\) convolutions, elongated \((5\times1)\) convolutions, elongated \((1\times5)\) convolutions, batch normalization, and ReLU nonlinearities. The \((5\times1)\) kernels learn slow-time Doppler structures and clutter trajectories, whereas the \((1\times5)\) kernels learn range-domain correlations and delay-spread clutter structures. Let \(\mathbf{F}\) denote an intermediate feature map. An anisotropic block computes
$\mathbf{F}_{\mathrm{st}}=
\mathrm{Conv}_{5\times1}(\mathbf{F})$ and $\mathbf{F}_{\mathrm{rg}}=
\mathrm{Conv}_{1\times5}(\mathbf{F})$,
followed by nonlinear feature fusion. This directional processing better distinguishes
between target signatures and clutter structures that exhibit different
geometric behavior across delay and slow-time dimensions.

\subsubsection{Attention-Gated Skip Connections}

The decoder incorporates attention-gated skip connections to suppress
clutter-dominated encoder activations before feature fusion. Let
\(\mathbf{S}\)
denote the encoder skip feature map and
\(\mathbf{G}\)
denote the decoder gating feature map. The attention mask is computed as $
\mathbf{A}
=
\sigma
\left(
\psi
\left(
\phi(\mathbf{S})
+
\theta(\mathbf{G})
\right)
\right)$,
where
\(\phi(\cdot)\),
\(\theta(\cdot)\),
and
\(\psi(\cdot)\)
are learned convolutional transformations and
\(\sigma(\cdot)\)
is the sigmoid activation function. The refined skip feature becomes
$
\mathbf{S}_{\mathrm{att}}
=
\mathbf{A}
\odot
\mathbf{S}$,
where
\(\odot\)
denotes elementwise multiplication. This mechanism suppresses clutter-dominated feature activations while preserving target-dominant spatial structures by emphasizing features consistent with the decoder's target-relevant context.

\subsubsection{Decoder Structure}

The decoder reconstructs the target-enhanced residual map through bilinear upsampling and feature concatenation with the attention-refined skip features. Each decoder stage employs anisotropic convolutional blocks similar to those used in the encoder. Anisotropic convolutions employ direction-dependent convolution kernels with unequal spatial support across dimensions. Kernels of size \(5\times1\) and \(1\times5\) are used to separately model slow-time Doppler structures and delay-domain correlations, respectively. The final output layer generates a two-channel estimate of the target residual:
\begin{equation}
\hat{\mathbf{Y}}
=
[
\hat{\mathbf{C}}_{\mathrm{target}}^{(\mathrm{I})},
\hat{\mathbf{C}}_{\mathrm{target}}^{(\mathrm{Q})}
].
\end{equation}

\subsubsection{Loss Function}

The network is trained using a combination of reconstruction and sparsity
losses:
\begin{align}
\mathcal{L}_{\mathrm{Attn}}
&=
\left\|
\hat{\mathbf{C}}_{\mathrm{target}}
-
\mathbf{C}_{\mathrm{target}}
\right\|_2^2
+
\lambda_s
\left\|
\hat{\mathbf{C}}_{\mathrm{target}}
\right\|_1
\nonumber\\
&\quad+
\lambda_{\mathrm{TV}}
\mathrm{TV}
(
\hat{\mathbf{C}}_{\mathrm{target}}
),
\end{align}
where
\(\lambda_s\)
controls sparsity regularization and
\(\lambda_{\mathrm{TV}}\)
controls total-variation regularization. The total-variation term promotes spatial smoothness while preserving
localized target peaks.

\subsubsection{Optimization and Training}

Training is performed using the AdamW optimizer with mini-batch gradient descent. Gradient clipping and weight decay are applied to improve optimization stability and generalization performance. After training, the network output is directly processed using CAF-based range-Doppler estimation and bistatic localization.

\section{Performance Evaluation}

\begin{table}[t]
\centering
\caption{Per-receiver bistatic range and radial-velocity estimation results.}
\label{tab:per_rx_results}
{\scriptsize
\setlength{\tabcolsep}{4pt}
\begin{tabular}{llrrrrrr}
\hline
Method & RX & $B_{\rm true}$ & $B_{\rm est}$ & $B_{\rm err}$ & $v_{\rm true}$ & $v_{\rm est}$ & $v_{\rm err}$ \\
\hline
NCS  & 1 & 331.16 & 293.11 & -38.05  & -25.61 & -2.17  & 23.44 \\
NCS  & 2 & 422.90 & 292.80 & -130.09 & -13.86 & 5.82   & 19.69 \\
NCS  & 3 & 533.39 & 312.32 & -221.07 & -49.05 & 6.85   & 55.91 \\
\hline
RMHP & 1 & 331.16 & 331.75 & 0.58    & -25.61 & -25.61 & -0.00 \\
RMHP & 2 & 422.90 & 429.64 & 6.74    & -13.86 & -13.88 & -0.01 \\
RMHP & 3 & 533.39 & 312.53 & -220.85 & -49.05 & 7.04   & 56.10 \\
\hline
DLCS-1 & 1 & 331.16 & 331.85 & 0.68   & -25.61 & -25.61 & -0.00 \\
DLCS-1 & 2 & 422.90 & 429.51 & 6.61   & -13.86 & -13.86 & 0.00 \\
DLCS-1 & 3 & 533.39 & 527.51 & -5.88  & -49.05 & -46.70 & 2.36 \\
\hline
\end{tabular}
}
\end{table}

\begin{table}[t]
\centering
\scriptsize
\caption{Localization and velocity estimation errors.}
\label{tab:summary_errors}
\begin{tabular}{lcccc}
\hline
Method
& \makecell{RMS $B$\\[-2pt]Error (m)}
& \makecell{Position\\[-2pt]Error (m)}
& \makecell{Speed\\[-2pt]Error (m/s)}
& \makecell{Angle\\[-2pt]Error (deg)} \\
\hline
NCS    & 149.715 & 114.985 & 22.184 & 114.065 \\
RMHP   & 127.570 & 158.673 & 9.509  & 156.948 \\
DLCS-1 & 5.122   & 4.462   & 0.445  & 0.045 \\
\hline
\end{tabular}
\end{table}

\begin{figure}[t]
\centering
\begin{subfigure}[t]{0.32\linewidth}
    \centering
    \includegraphics[width=\linewidth]{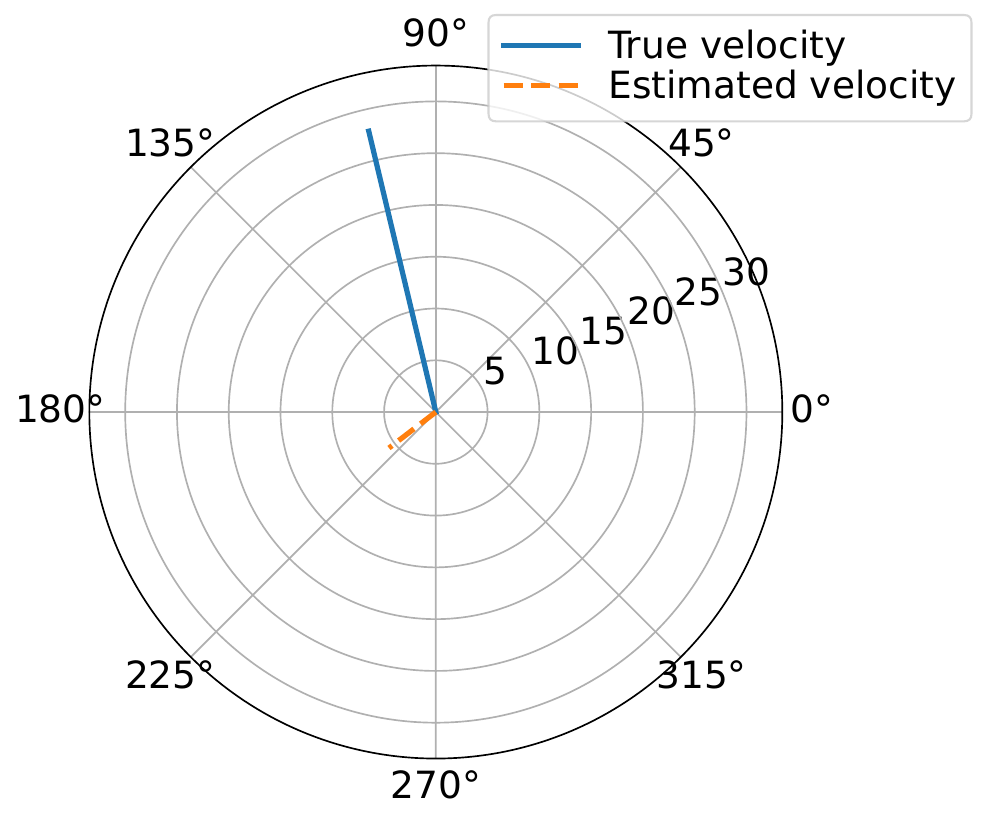}
    \caption{NCS}
\end{subfigure}
\hfill
\begin{subfigure}[t]{0.32\linewidth}
    \centering
    \includegraphics[width=\linewidth]{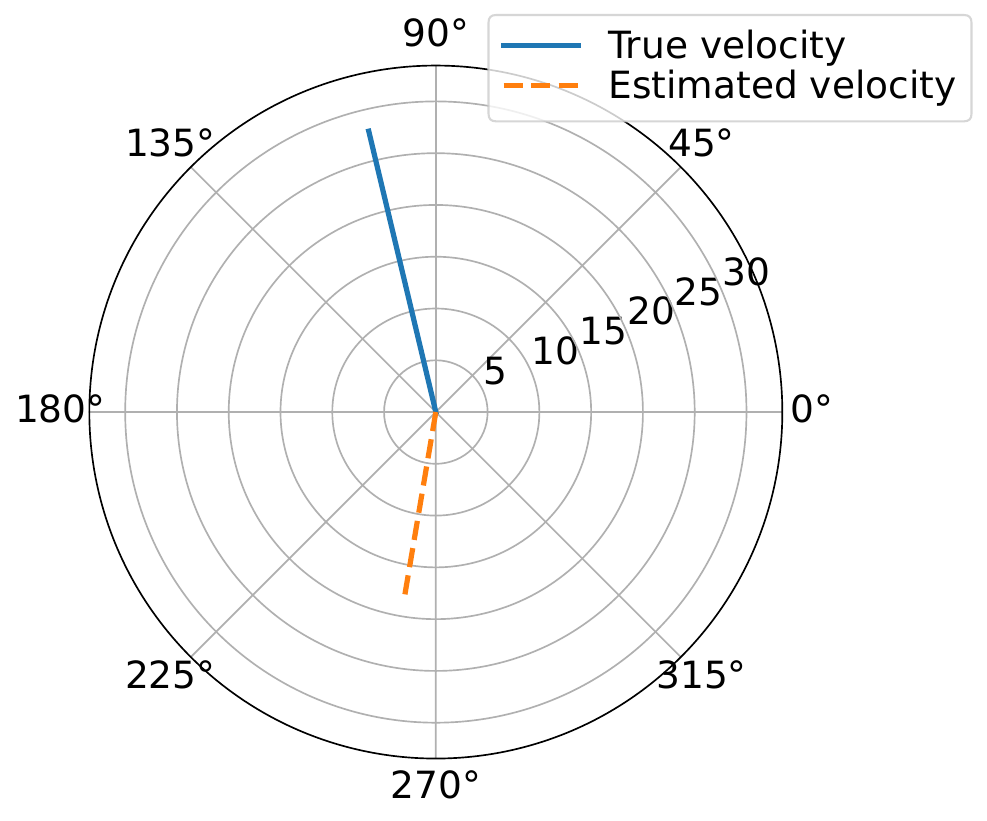}
    \caption{RMHP}
\end{subfigure}
\hfill
\begin{subfigure}[t]{0.32\linewidth}
    \centering
    \includegraphics[width=\linewidth]{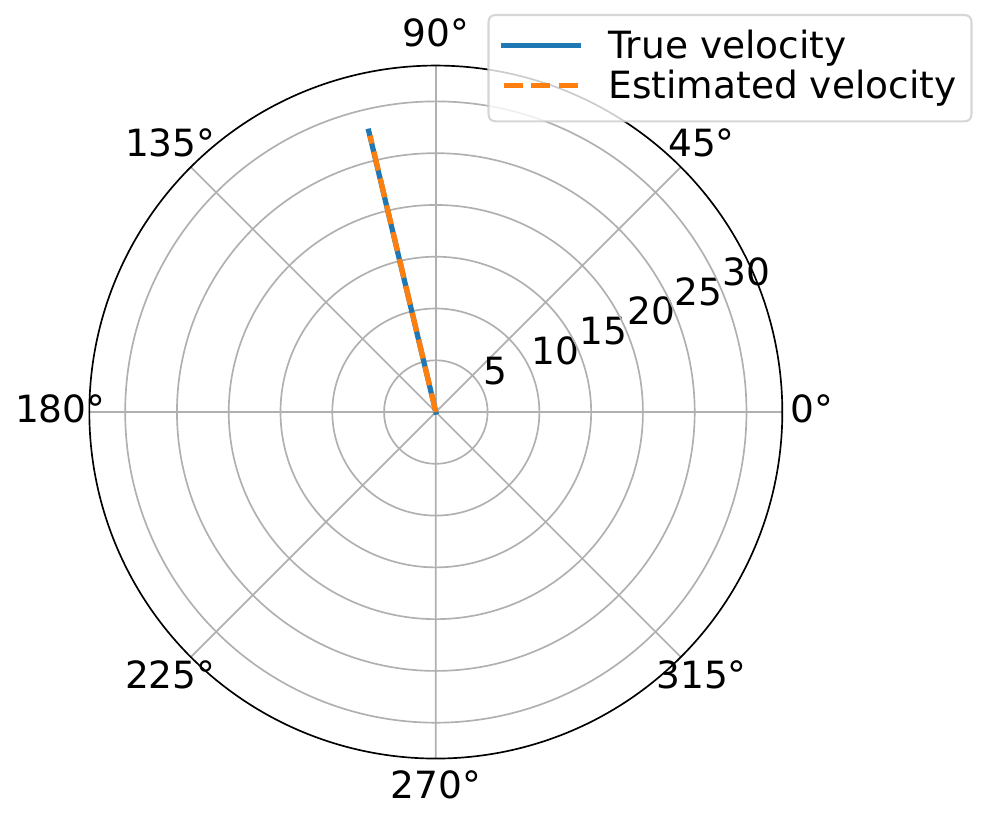}
    \caption{DLCS-1}
\end{subfigure}
\caption{Velocity estimation results.}

\label{fig:velocity_results}
\end{figure}

\begin{figure}[t]
\centering
\begin{subfigure}[t]{0.32\linewidth}
    \centering
    \includegraphics[width=\linewidth]
    {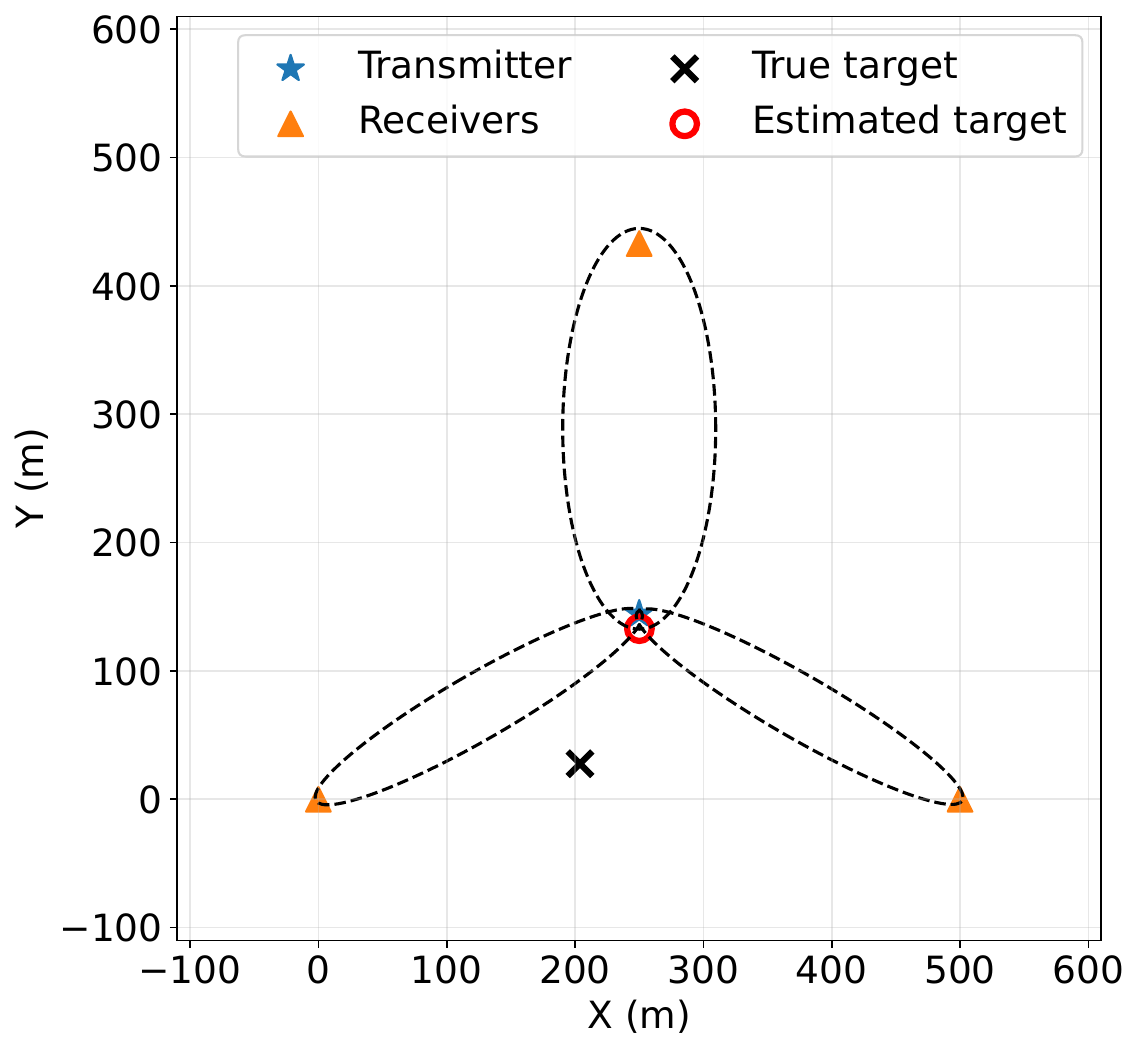}
    \caption{NCS}
\end{subfigure}
\hfill
\begin{subfigure}[t]{0.32\linewidth}
    \centering
    \includegraphics[width=\linewidth]
    {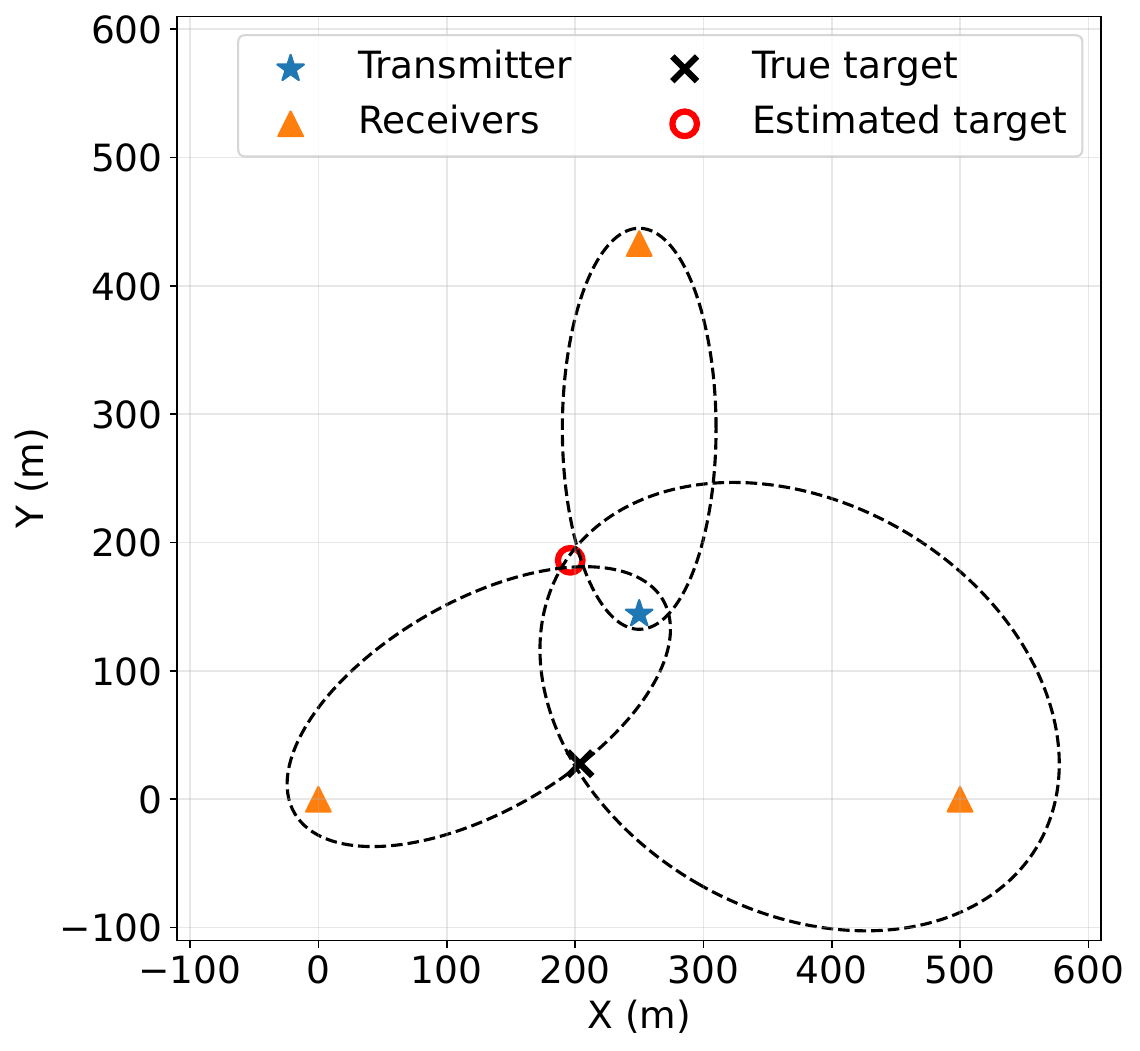}
    \caption{RMHP}
\end{subfigure}
\hfill
\begin{subfigure}[t]{0.32\linewidth}
    \centering
    \includegraphics[width=\linewidth]
    {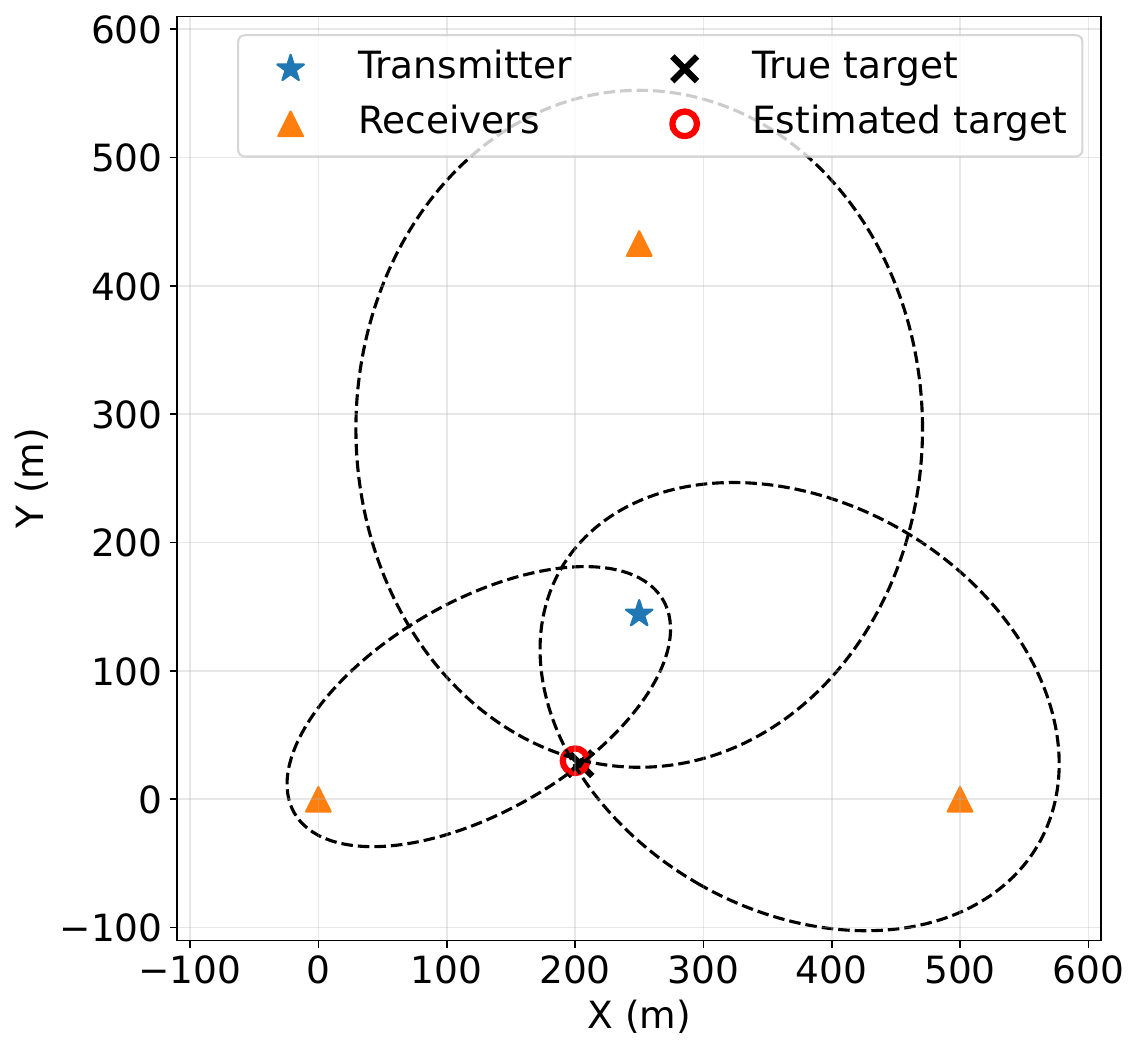}
    \caption{DLCS-1}
\end{subfigure}
\caption{Target localization results.}
\vspace{-0.4cm}
\label{fig:position_results}
\end{figure}

\begin{figure}[t]
\centering
\begin{subfigure}[t]{\linewidth}
    \centering
    \includegraphics[width=\linewidth]{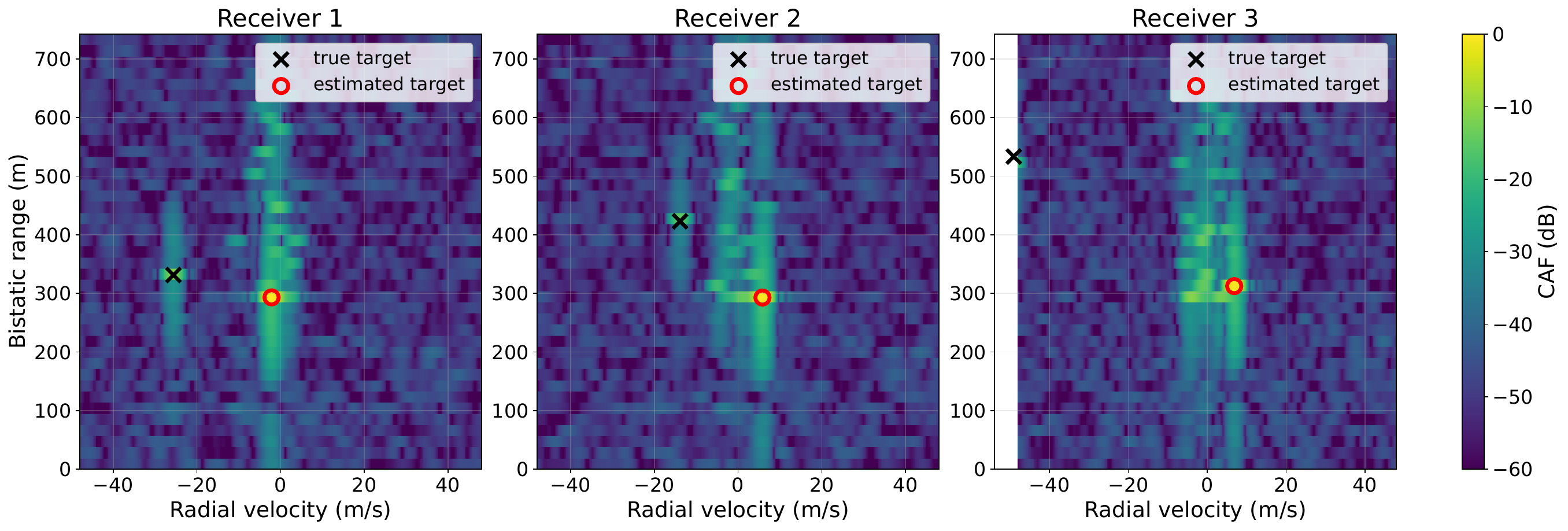}
    \caption{NCS}
\end{subfigure}

\vspace{1mm}

\begin{subfigure}[t]{\linewidth}
    \centering
    \includegraphics[width=\linewidth]{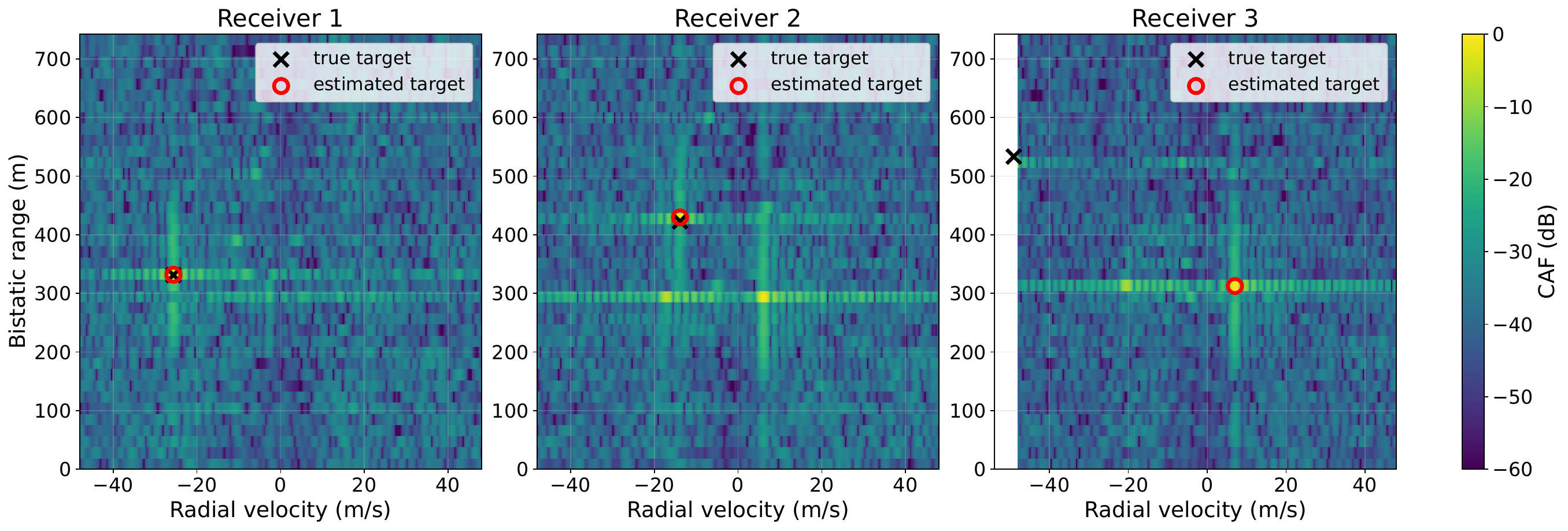}
    \caption{RMHP}
\end{subfigure}

\vspace{1mm}

\begin{subfigure}[t]{\linewidth}
    \centering
    \includegraphics[width=\linewidth]{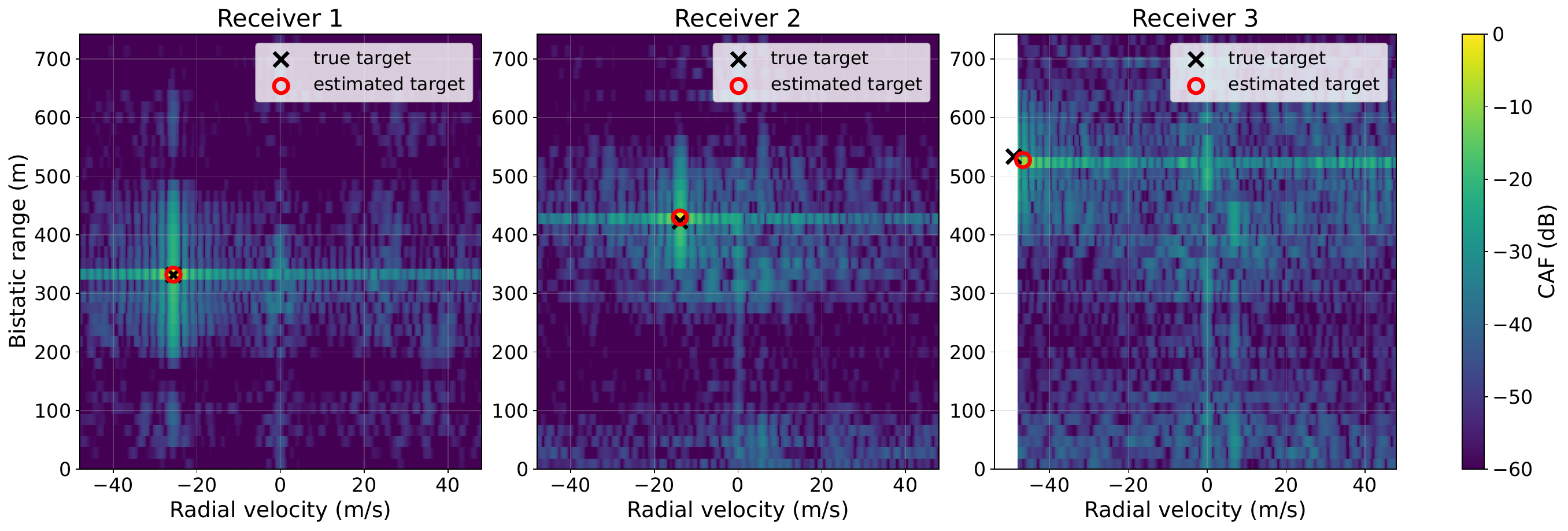}
    \caption{DLCS-1}
\end{subfigure}
\caption{Bistatic range-Doppler maps.}
\label{fig:range_doppler_results}
\end{figure}

For the numerical results, we assume a transmit power of 40~dBm, transmit and receive antenna gains of 10~dBi, a target RCS of 4~\(\mathrm{m}^2\), and a noise variance of \(10^{-3}\). All DL models were trained using NVIDIA RTX PRO 6000 Blackwell GPUs. The DLCS-1 and DLCS-2 architectures contain 1,885,186 and 12,398,789 trainable parameters, respectively. Let \(N_{\mathrm{train}}\) and \(N_{\mathrm{test}}\) denote the numbers of training and test samples, respectively, and let \(N_{\mathrm{epoch}}\) denote the number of training epochs. For each sample, the transmitter and receiver geometry, target position and velocity, clutter scatterer locations, and additive noise realizations are randomly generated. Clutter effects are controlled by the values of \(N_c\), \(\alpha_c\), and \(\sigma_{f,c}\). The training and test data are generated independently, with clutter density, strength, and Doppler spread randomized during training.

We first consider a moderate training setting of \(N_{\mathrm{train}}\!=\!500\) and \(N_{\mathrm{epoch}}\!=\!200\) under a strong clutter environment with \(N_c\!=\!40\), \(\alpha_c\!=\!1\), and \(\sigma_{f,c}\!=\!30\)~Hz. For a randomly selected test sample (\(N_{\mathrm{test}}\!=\!1\)), Tables~\ref{tab:per_rx_results} and~\ref{tab:summary_errors} compare DLCS-1 with NCS and RMHP. NCS produces severe bistatic-range and radial-velocity estimation errors, leading to large localization and velocity estimation errors, as shown in Figs.~\ref{fig:velocity_results} and~\ref{fig:position_results}. RMHP, the strongest conventional baseline, improves the estimates by suppressing slowly varying clutter, but fails to accurately recover the target response for the other receivers, causing large localization and velocity-estimation errors. DLCS-1 accurately estimates bistatic range and radial velocity for all receivers, achieving a position error of 4.462~m and speed and angle estimation errors of only 0.445~m/s and \(0.045^\circ\), respectively. 

The range-Doppler maps in Fig.~\ref{fig:range_doppler_results} further illustrate these performance differences. For NCS and RMHP, clutter obscures the target response, causing large estimation errors. In contrast, DLCS-1 produces cleaner range-Doppler maps with localized target peaks closely aligned with the true target locations. As a result, DLCS-1 achieves highly accurate velocity estimation (Fig.~\ref{fig:velocity_results}) and target localization (Fig.~\ref{fig:position_results}).

\begin{table*}[t]
\centering
\scriptsize
\setlength{\tabcolsep}{4.8pt}
\caption{Normalized success rate $\eta_{\text{succ}}$ for different clutter-suppression methods.}
\label{tab:normalized_success_rates}
\begin{tabular}{ccc|ccccccccccccc}
\hline
$N_c$ &
$\alpha_c$ &
$\sigma_{f,c}$ &
NCS &
SP-AD &
SP-R1 &
DPSS &
CMTD &
TVHP &
RMHP &
WHP &
ALMS &
POLYHP &
DLCS-1 \\
\hline

40 & 2 & 30 Hz
& 1.72\%
& 24.30\%
& 1.72\%
& 2.62\%
& 1.72\%
& 18.41\%
& 25.70\%
& 10.56\%
& 10.23\%
& 2.62\%
& 62.77\% \\

40 & 1 & 30 Hz
& 2.86\%
& 35.27\%
& 3.03\%
& 5.16\%
& 3.52\%
& 29.46\%
& 35.11\%
& 18.74\%
& 17.92\%
& 5.65\%
& 74.39\% \\

20 & 2 & 30 Hz
& 2.85\%
& 36.92\%
& 3.83\%
& 5.62\%
& 3.26\%
& 28.61\%
& 40.18\%
& 22.00\%
& 19.89\%
& 5.62\%
& 72.21\%\\

20 & 1 & 30 Hz
& 5.46\%
& 46.70\%
& 6.68\%
& 10.35\%
& 6.03\%
& 38.30\%
& 48.17\%
& 32.19\%
& 29.99\%
& 10.51\%
& 80.93\% \\

40 & 1 & 10 Hz
& 2.29\%
& 65.90\%
& 5.40\%
& 29.84\%
& 4.09\%
& 59.93\%
& 85.61\%
& 54.62\%
& 46.85\%
& 2.62\%
& 90.27\% \\

20 & 1 & 10 Hz
& 4.97\%
& 73.94\%
& 11.64\%
& 46.42\%
& 9.77\%
& 65.55\%
& 89.98\%
& 68.08\%
& 62.87\%
& 43.08\%
& 94.22\% \\

\hline
\end{tabular}
\end{table*}

\begin{table}[t]
\centering
\scriptsize
\caption{Normalized success rate comparison of DLCS-1 and DLCS-2.}
\label{tab:table_iii_dlcs_improved_comparison}
\begin{tabular}{ccc cc cc}
\hline
$N_c$ &
$\alpha_c$ &
$\sigma_{f,c}$ &
$N_{\mathrm{train}}$ &
$N_{\mathrm{epoch}}$ &
DLCS-1 &
DLCS-2\\
\hline

\multirow{2}{*}{40} &
\multirow{2}{*}{2} &
\multirow{2}{*}{30 Hz}
& 500  & 200 & $62.77\%$ & $66.61\%$ \\
& & & 5000 & 500 & $71.19\%$ & $72.42 \%$ \\
\hline
\multirow{2}{*}{40} &
\multirow{2}{*}{1} &
\multirow{2}{*}{30 Hz}
& 500  & 200 & $74.39\%$ & $78.48\%$ \\
& & & 5000 & 500 & $80.44\%$ & $84.29\%$ \\
\hline
\multirow{2}{*}{20} &
\multirow{2}{*}{2} &
\multirow{2}{*}{30 Hz}
& 500  & 200 & $72.21\%$ & $74.16\%$ \\
& & & 5000 & 500 & $76.69\%$ & $78.24\%$ \\
\hline
\multirow{2}{*}{20} &
\multirow{2}{*}{1} &
\multirow{2}{*}{30 Hz}
& 500  & 200 & $80.93\%$ & $88.18\%$ \\
& & & 5000 & 500 & $87.61 \%$ & $92.42\%$ \\

\hline
\multirow{2}{*}{40} &
\multirow{2}{*}{1} &
\multirow{2}{*}{10 Hz}
& 500  & 200 & $90.27\%$ & $90.52\%$ \\
& & & 5000 & 500 & $95.59\%$ & $ 95.67\%$ \\

\hline
\multirow{2}{*}{20} &
\multirow{2}{*}{1} &
\multirow{2}{*}{10 Hz}
& 500  & 200 & $94.22\%$ & $98.05\%$ \\
& & & 5000 & 500 & $98.70\%$ & $99.41\%$ \\

\hline
\end{tabular}
\end{table}

Next, we increase the number of test samples to \(N_{\mathrm{test}}=2000\) and statistically compare DLCS-1 with conventional clutter-suppression baselines under different clutter settings specified by \(N_c\), \(\alpha_c\), and \(\sigma_{f,c}\). For each clutter case, the success rate is defined as the percentage of trials with position error below \(T=10\)~m, and \(\eta_{\text{succ}}\) is computed as the normalized success rate relative to the reference method CUR of Sec.~\ref{sec:CUR}.

As shown in Table~\ref{tab:normalized_success_rates}, DLCS-1 consistently achieves a higher normalized success rate \(\eta_{\text{succ}}\) than all non-reference clutter-suppression methods across all clutter environments. Conventional methods degrade significantly under dense and dynamic clutter. RMHP is the strongest conventional baseline, yet remains substantially outperformed by DLCS-1.

Table~\ref{tab:table_iii_dlcs_improved_comparison} compares \(\eta_{\text{succ}}\) of DLCS-1 and DLCS-2 for different clutter and training configurations. As \(N_{\mathrm{train}}\) and \(N_{\mathrm{epoch}}\) increase to 5000 and 500, respectively, DLCS-2 approaches the performance of the reference method while consistently outperforming DLCS-1 across all clutter settings.

\begin{figure}[t]
\centering

\begin{subfigure}{0.49\columnwidth}
\centering
\includegraphics[width=0.95\linewidth]{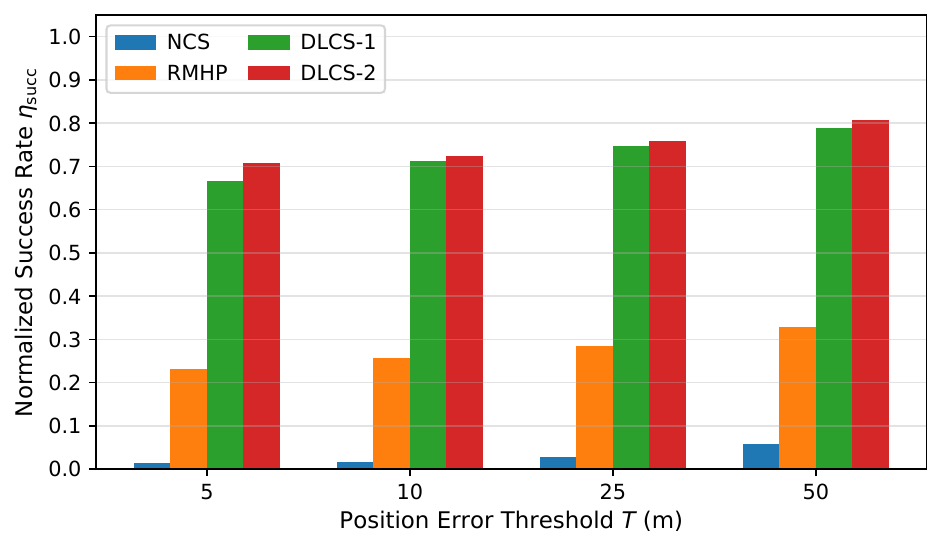}
\caption{$N_c=40, \alpha_c=2, \sigma_{f,c}=30$~Hz.}
\vspace{0.1cm}
\end{subfigure}
\hfill
\begin{subfigure}{0.49\columnwidth}
\centering
\includegraphics[width=0.95\linewidth]{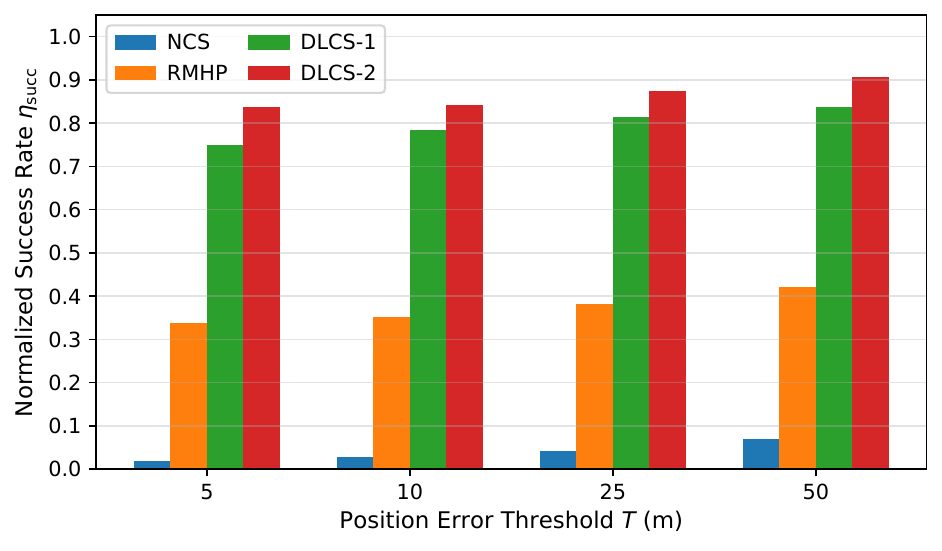}
\caption{$N_c=40, \alpha_c=1, \sigma_{f,c}=30$~Hz.}
\vspace{0.1cm}
\end{subfigure}

\begin{subfigure}{0.49\columnwidth}
\centering
\vspace{0.1cm}
\includegraphics[width=0.95\linewidth]{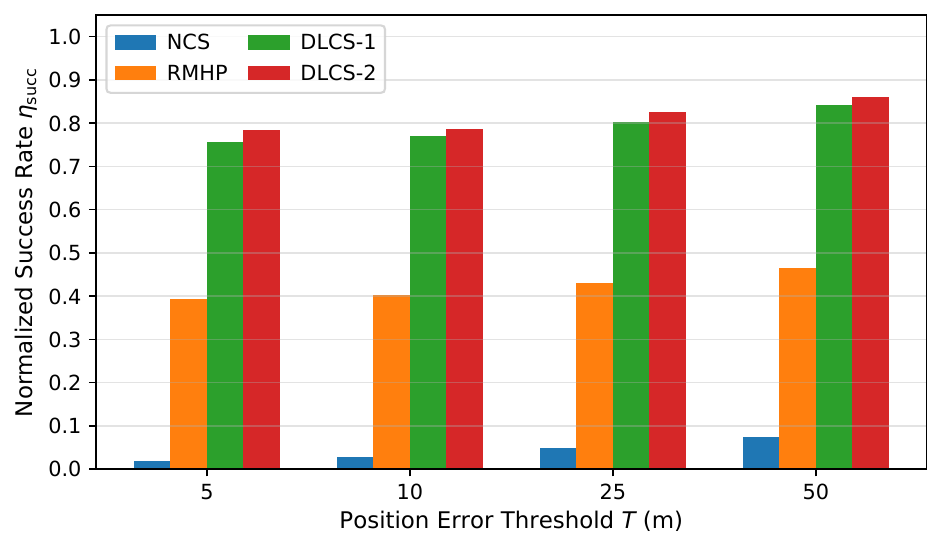}
\caption{$N_c=20, \alpha_c=2, \sigma_{f,c}=30$~Hz.}
\vspace{0.1cm}
\end{subfigure}
\hfill
\begin{subfigure}{0.49\columnwidth}
\centering
\vspace{0.1cm}
\includegraphics[width=0.95\linewidth]{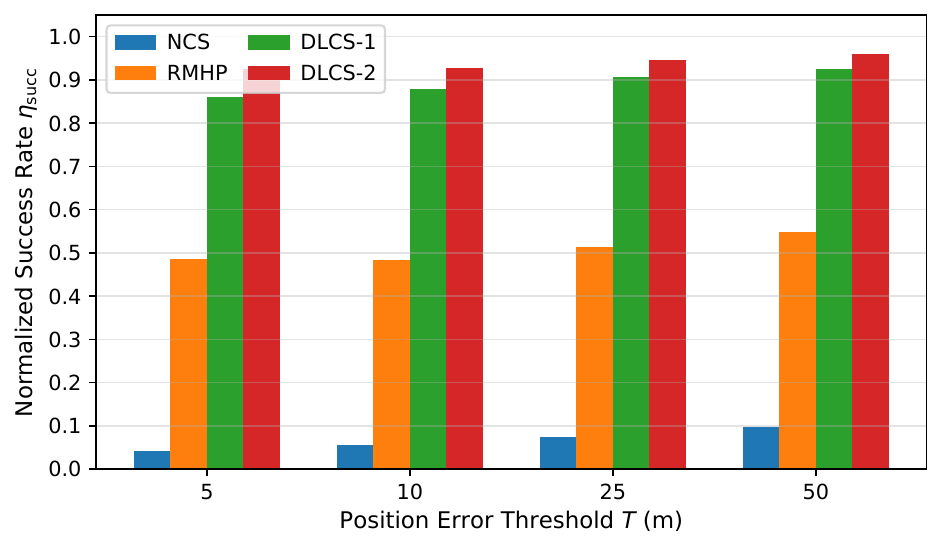}
\caption{$N_c=20, \alpha_c=1, \sigma_{f,c}=30$~Hz.}
\vspace{0.1cm}
\end{subfigure}

\begin{subfigure}{0.49\columnwidth}
\centering
\vspace{0.1cm}
\includegraphics[width=0.95\linewidth]{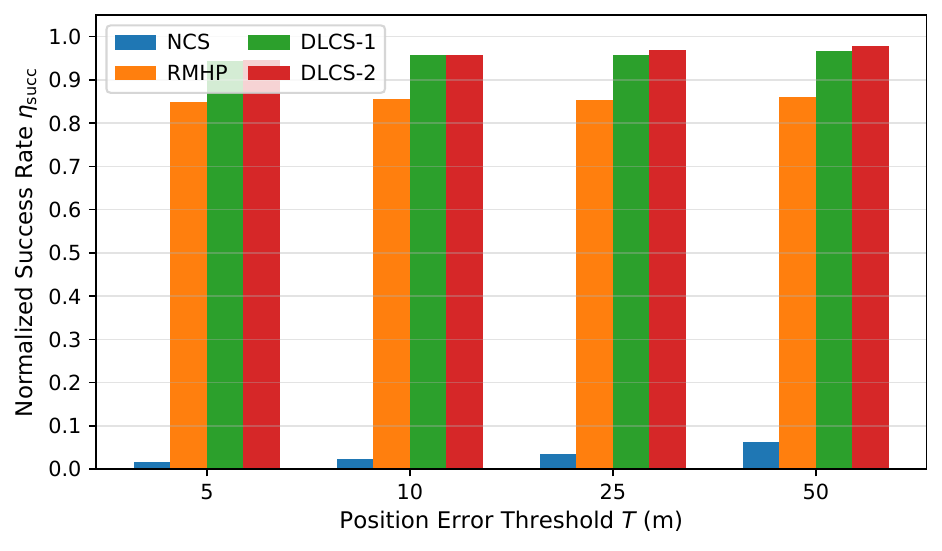}
\caption{$N_c=40, \alpha_c=1, \sigma_{f,c}=10$~Hz.}
\vspace{0.1cm}
\end{subfigure}
\hfill
\begin{subfigure}{0.49\columnwidth}
\centering
\vspace{0.1cm}
\includegraphics[width=0.95\linewidth]{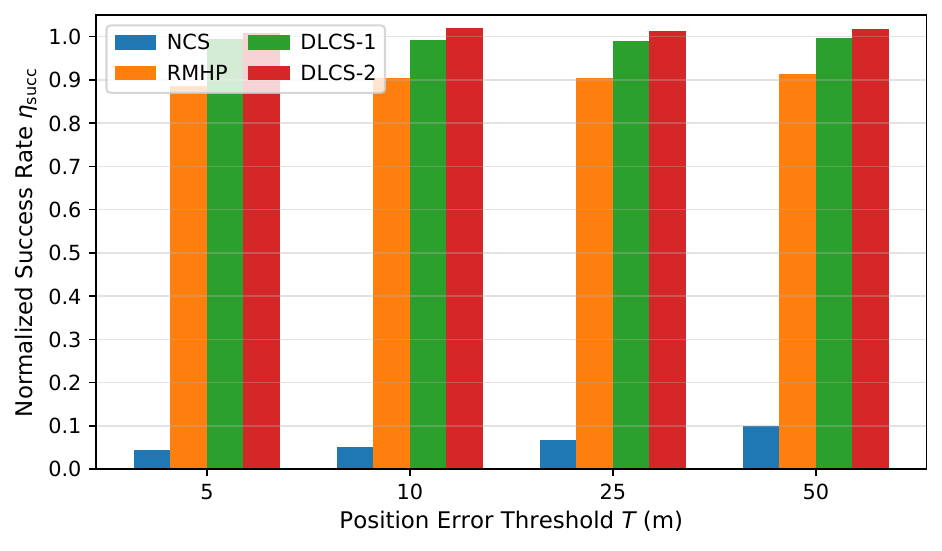}
\caption{$N_c=20, \alpha_c=1, \sigma_{f,c}=10$~Hz.}
\vspace{0.1cm}
\end{subfigure}

\caption{Normalized success rate $\eta_{\text{succ}}$ versus position error threshold $T$ (m).}
\label{fig:success_rate_comparison}

\end{figure}

Fig.~\ref{fig:success_rate_comparison} shows \(\eta_{\text{succ}}\)  as a function of position error threshold \(T\). Across all \(T\) and clutter settings, the DLCS methods, trained with \(N_{\mathrm{train}}=5000\) and \(N_{\mathrm{epoch}}=500\), consistently outperform NCS and RMHP. DLCS-2 achieves the best performance, followed closely by DLCS-1, demonstrating the effectiveness of DL-based clutter suppression in multistatic ISAC.

\section{Conclusion \label{sec:conclusion}}

In this paper, we studied clutter suppression in multistatic ISAC using 5G NR PRS for bistatic range-Doppler estimation, target localization, and velocity estimation in dense and dynamic clutter environments. We showed that clutter can severely degrade sensing accuracy of multistatic ISAC and that conventional signal processing methods for clutter suppression provide only limited gains under challenging clutter conditions. To improve robustness of multistatic ISAC against clutter, we introduced DL-based clutter suppression methods to directly enhance target responses from clutter-corrupted observations. These methods substantially improve clutter suppression and reduce sensing errors relative to conventional approaches across diverse clutter conditions in multistatic ISAC.

\bibliographystyle{IEEEtran}
\bibliography{references}
\end{document}